**Preventing the reconstruction of the polar discontinuity at oxide heterointerfaces**

By *H. Boschker, J. Verbeeck, R. Egoavil, S. Bals, G. van Tendeloo, M. Huijben, E.P. Houwman, G. Koster, D.H.A. Blank,* and *G. Rijnders**

[*]     Dr. H. Boschker, Dr. M. Huijben, Dr. E.P. Houwman, Dr. G. Koster, Prof. D.H.A. Blank, Prof. G. Rijnders
        Faculty of Science and Technology and MESA+ Institute for Nanotechnology
        University of Twente
        7500 AE, Enschede, The Netherlands
        E-mail: a.j.h.m.rijnders@utwente.nl
        Dr. J. Verbeeck, R. Egoavil, Dr. S. Bals, Prof. G. van Tendeloo
        Electron Microscopy for Materials Science (EMAT)
        University of Antwerp
        2020 Antwerp, Belgium

Keyword: interface engineering, oxide heterostructure, polar discontinuity, epitaxy, perovskite

Perovskite oxide heteroepitaxy receives much attention because of the possibility to combine the diverse functionalities of perovskite oxide building blocks. A general boundary condition for the epitaxy is the presence of polar discontinuities at heterointerfaces. These polar discontinuities result in reconstructions, often creating new functionalities at the interface. However, for a significant number of materials these reconstructions are unwanted as they alter the intrinsic materials properties at the interface. Therefore, a strategy to eliminate this reconstruction of the polar discontinuity at the interfaces is required. We show that the use of compositional interface engineering can prevent the reconstruction at the $La_{0.67}Sr_{0.33}MnO_3/SrTiO_3$ (LSMO/STO) interface. The polar discontinuity at this interface can be removed by the insertion of a single $La_{0.33}Sr_{0.67}O$ layer, resulting in improved interface magnetization and electrical conductivity.

## 1. Introduction

The properties of interfaces between the ferromagnetic metallic LSMO and the insulating STO have been well studied, because of the relevance in various devices, such as diodes[1,2] transistors[3] and magnetic tunnel junctions.[4] Although working devices have been reported, it is well known that the properties of the LSMO are deteriorating at the interface,[5] resulting in an interfacial dead layer for both the conductivity and the magnetization.[6] This reduction is attributed to either a valence change at the interface,[7] a change in orbital ordering[8] or a structural change.[9,10] In several studies, use is made of interface engineering to improve the devices. E.g., a 2 unit cell LaMnO$_3$ layer was inserted at the interface to compensate for an observed valence change[11,12] and a single SrO layer was inserted to modify the Schottky barrier height of the devices.[2,3] These modifications, however, do not remove the polar discontinuity and therefore the driving force for reconstructions is not eliminated.

Two different atomic stacking sequences are possible for the LSMO/STO interfaces, because of the possible $A$O or $B$O$_2$ termination of each material (with respect to the model perovskite $AB$O$_3$). In case of LSMO, this leads to a La$_{0.67}$Sr$_{0.33}$O or MnO$_2$ termination. The La$_{0.67}$Sr$_{0.33}$O terminated interface is shown in Fig. 1a. At this interface the atomic stacking sequence is SrO-TiO$_2$-La$_{0.67}$Sr$_{0.33}$O-MnO$_2$. The MnO$_2$ terminated interface is shown in Fig. 1b. Here the atomic stacking sequence is TiO$_2$-SrO-MnO$_2$-La$_{0.67}$Sr$_{0.33}$O. The net charge per area unit in the sub unit cell layers in the STO is 0, while it is alternatingly 2/3$e$ and -2/3$e$ in the LSMO. Here $e$ is the electron charge. Both interface configurations are polar, and without reconstruction would result in a diverging electrostatic potential, as indicated in the figures. As LSMO is conducting, the mobile charges can screen the diverging potential, resulting in a compensating charge transfer of 1/3$e$, as indicated in the figures. This screening, however, is only partial due to the fact that the charges are confined to the Mn sublattice. In Fig. 1a and 1b the electrostatic potential after the screening (effectively an electronic reconstruction of the polar discontinuity) is shown as well. Here it is assumed that the screening occurs completely in the first $MnO_2$ layer, which is consistent with the Thomas Fermi length in LSMO of ≈ 0.31 nm.[2] The diverging potential is eliminated, but a band offset, especially in the case of the $La_{0.67}Sr_{0.33}O$ terminated interface, is present and can be expected to result in intermixing of the cations.[13] Note that the $Ti^{4+}$ valence state is stable at all these interfaces, as the Ti $t_{2g}$ levels are 1 eV higher in energy compared to the Mn $e_g$ levels.[7]

In order to remove the polar discontinuity, a sub unit cell atomic layer with a net charge of $1/3e$ per unit cell has to be inserted at the interface. This is shown in Fig. 1c and 1d, in which two methods are shown for interface engineering by the insertion of a single atomic layer of $La_{0.33}Sr_{0.67}O$ through the deposition of either a $La_{0.33}Sr_{0.67}MnO_3$ layer (I) or a $La_{0.33}Sr_{0.67}TiO_3$ layer (II). This gives an atomic stacking sequence at the interface of $SrO$-$TiO_2$-$La_{0.33}Sr_{0.67}O$-$MnO_2$-$La_{0.67}Sr_{0.33}O$. In this case, the diverging potential is absent and no driving force for reconstruction exists. A band offset is present at this interface as well, so some intermixing can still be expected.

Here, we present experimental results of thin LSMO films with the different interface configurations. The non polar interface as shown in Fig. 1c can be realized with the use of compositional interface engineering. Using scanning transmission electron microscopy (STEM), it is shown that the samples with the polar discontinuity are intermixed over a distance of 2 unit cells (uc). In contrast, samples with engineered interfaces, and therefore without the polar discontinuity, have a sharp interface. The latter samples have improved magnetization and electrical conductivity compared to the non interface engineered heterostructures.

## 2. Results and Discussion

Heterostructures with and without interface engineering (denoted IE and nIE, respectively) were grown by pulsed laser deposition (PLD) at 800 °C in an oxygen environment of 0.27 mbar, as described in detail in the experimental section. The structures consist of a $TiO_2$ terminated STO substrate, a variable thickness LSMO layer and an STO capping layer with a thickness of 5 unit cells. In order to realize the interface engineered heterostructure, first a single unit cell layer of $La_{0.33}Sr_{0.67}MnO_3$ is grown. This is followed by the growth of $n$-1 layers of LSMO, then a single unit cell of $La_{0.33}Sr_{0.67}TiO_3$ and finally, 4 unit cell layers of STO. This recipe results in heterostructures with $n$ layers of Mn ions, which can therefore be compared to the $n$ unit cell layer LSMO/STO structure without the interface engineering.

Figures 2a and 2b show the RHEED specular spot intensities during the depositions of respectively a non interface engineered heterostructure and an interface engineered heterostructure. The clear oscillations of the intensity indicate layer-by-layer growth and allow for the possibility to grow single unit cell layers. After the growth of each layer, the RHEED diffraction images showed clear two dimensional spots, indicating diffraction from smooth crystalline surfaces. The full width at half maximum (FWHM) of the specular spot intensity was also monitored along the $(10)_{pc}$ direction (not shown). The FWHM after growth of each LSMO and STO unit cell layer was identical to the initial smooth step and terrace STO surface before growth confirming the layer-by-layer growth mode. Figures 2c and 2d show atomic force microscopy (AFM) measurements of the smooth surface topography of respectively the non interface engineered (nIE) and interface engineered (IE) samples exhibiting the unit cell height terrace steps, similar to the surface of the original STO substrate. This suggests a limitation to the roughness of the interfaces in the heterostructures of less than a unit cell.

Two dimensional spectrum-images were acquired with atomic resolution scanning transmission electron microscopy combined with electron energy loss spectroscopy (STEM-EELS) to investigate the interface structure for both nIE and IE samples. To improve the signal to noise ratio in the EELS spectra, principal components analysis (PCA) was applied. This analysis technique is a powerful tool to reduce the noise from STEM-EELS data which enables one to extract the essential chemical information.[14,15] Quantitative elemental maps corresponding to La $M_{4,5}$, Mn $L_{2,3}$, Ti $L_{2,3}$ and O K edges (not shown) are generated by applying PCA. Figure 3 shows the quantitative maps for the nIE (top left) sample and the IE (bottom left) sample, with La (red), Mn (green) and Ti (blue). We concentrate on the bottom (substrate/film) interface in both samples because of unwanted effects of electron beam damage near the top interface during the STEM-EELS acquisition. At this interface, a different shape for the La concentration profile is observed between the two samples. The profile is wide for the nIE sample and sharp for the IE sample. No remarkable changes in the fine structure (not shown) are found for the La $M_{4,5}$, Mn $L_{2,3}$, Ti $L_{2,3}$ and O K edges acquired at the interface in comparison to the ones acquired further away from the interface.

The middle panels of Fig. 3 show quantitative elemental profiles for La, Mn, Ti and O obtained from an integrated signal region across the STO/LSMO/STO stack. The right panels of Fig. 3 show a comparison between the experimental La profile and a simple weighted Gaussian model, which is based on an ideal Gaussian probe shape convoluted with the discrete La lattice occupancies in the $La_{0.67}Sr_{0.33}O$ and $La_{0.33}Sr_{0.67}O$ terminated interface for nIE and IE. This comparison shows a high level of La diffusion into the STO layer for nIE and very low La diffusion for the IE sample. Indeed, a reasonable fit with the model is found for the IE sample, but a large deviation is observed for the nIE sample. This deviation for the nIE sample indicates a diffusion process that is not implemented in the model. The diffusion reaches zero at approximately 2 nm from the interface in the STO substrate.

Figure 4 presents temperature dependent magnetization and resistivity measurements of the samples. The magnetic field was applied in-plane along the $[100]_c$ STO crystal direction during the magnetization measurements. Figure 4a shows the temperature dependence of the saturation magnetization for the $n$ = 5, 6, 8 and 13 uc samples. A clear difference between the IE and nIE samples is observed. The IE samples have significantly higher saturation magnetization and Curie temperature ($T_C$). Details about the determination of the $T_C$ are given in the supplementary information.[16] Figure 4b presents the temperature dependence of the resistivity of the samples. For all thicknesses, the interface engineering results in a lower resistivity. The $n$ = 5 samples and the nIE $n$ = 6 sample are insulating at low temperature. The $n$ = 6 IE sample has an upturn in the resistivity at low temperature but conductivity was observed down to the lowest temperature (10 K) in the measurement. An enhancement of the magnetization and electrical conductivity can still be observed for thicker LSMO layers in the 13 uc samples, however, the effect is smaller as the enhanced interface properties provide a smaller contribution to the properties of the complete LSMO layer when thicker layers are studied. The obtained properties of the non interface engineered 13 uc sample are in good agreement with previously obtained results in ultrathin LSMO films.[6] There it was shown that LSMO layers with a thickness of 13 unit cells exhibit properties close to bulk values, but still contain a small deviation. Here, we demonstrate that additional interface engineering induces an enhancement in magnetization and electrical conductivity of a 13 uc sample to bring the properties very close to bulk values.

Figures 5a, 5b and 5c summarize the data of all the LSMO/STO heterostructures. For all thicknesses, interface engineering resulted in a higher $T_C$, a higher saturation magnetization and higher electrical conductivity. Typically, the improvement is 0.5 $\mu_B$/Mn in the saturation magnetization and 20 K in the $T_C$ for LSMO layers of 5 to 8 unit cells, while a smaller enhancement is observed for thicker layers. Even though we observe an improvement in magnetization and electrical conductivity because of the interface engineering, a dead layer is still present. LSMO's electrical dead layer was previously reported to be 8 unit cells.[6,17] For the nIE samples in this study, the electrical dead layer is 6 uc while for the IE samples it is 5 uc. The continued presence of the dead layer indicates that an additional mechanism still results in the deterioration of the intrinsic properties of the LSMO. Based on our study we can exclude the contribution of valence changes at the interface, which have not been observed in the fine structure of the Mn spectrum at the $L_{2,3}$ edge. The change in orbital occupation was also not observed in samples grown in our group.[6] Therefore, the local changes in crystal structure and the resulting changes in exchange coupling between Mn ions at the interface[9] is left as the most likely explanation for the dead layer in LSMO. The thickness of the dead layer is determined by the specific epitaxial heterostructure in which the LSMO layer is grown as demonstrated by a further reduction of the dead layer in superlattices.[18,19]

The main result of this research is the improvement of the properties by using interface engineering. The samples for which the polar discontinuities at the interfaces were artificially removed show less intermixing at the interfaces, especially the film-substrate interface, and improved magnetization and electrical conductivity compared to the non interface engineered samples. Therefore, the improvement in functional properties can be attributed to a reduction of the intermixing in the interface engineered samples. The intermixing results in larger chemical disorder at the interface and therefore larger local crystal structure variations. The magnetic exchange coupling between interfacial Mn ions is therefore reduced for the non interface engineered samples as it depends on the local crystal structure[9]. In general, intermixing is expected at polar interfaces as it reduces the electrostatic band offset.[13] Significant intermixing was only observed at the bottom interface of the non interface engineered sample, which corresponds to the $La_{0.67}Sr_{0.33}O$ terminated interface shown in Fig. 1a. This interface configuration has the largest band offset and therefore, the observations of the intermixing are in good agreement with the model. The expected valence changes of Mn at the $MnO_2$ and $La_{0.67}Sr_{0.33}O$ terminated interfaces, however, are not observed. This suggests that the polar discontinuity is not compensated by an electronic reconstruction, but by a structural reconstruction.

## 3. Conclusion

In conclusion, we have shown that with the insertion of a single $La_{0.33}Sr_{0.67}O$ atomic layer at the interface, the polar discontinuity is compensated and reconstruction is prevented. Measurements on ultrathin LSMO layers without polar discontinuity driven reconstructions showed an improvement in the functional properties. The reduction of properties in the non interface engineered samples can be attributed to a larger amount of intermixing in these samples compared to the interface engineered samples. This intermixing is the result of the structural reconstruction of the polar discontinuity. It is suggested that this method of interface engineering is also interesting for other mixed valence compounds, such as the cuprate superconductors. These compounds generally have a polar discontinuity at the interface with the substrate, possibly affecting film growth, and at the interface with other materials in the required device stack, with a possible reduction of the superconductivity at the interface.[20] Interface engineering is expected to result in improved growth and quality of the thin films and improved properties at the interfaces.

## 4. Experimental section

All heterostructures were fabricated on $TiO_2$ terminated[21] STO substrates obtained from Crystec GmbH and were grown by pulsed laser deposition (TSST system). The substrate temperature during growth was 750-800 °C in an oxygen environment of 0.27 mbar. The laser beam was produced by a 248-nm-wavelength KrF excimer laser (LPXProTM from Coherent, Inc.) with a typical pulse duration of 20-30 ns. With a 4 by 15 mm rectangular mask the most homogeneous part of the laser beam was selected. An image of the mask was created on the stoichiometric targets ($SrTiO_3$ (Single crystal from Crystec GmbH), $La_{0.33}Sr_{0.67}MnO_3$, $La_{0.67}Sr_{0.33}MnO_3$ and $La_{0.33}Sr_{0.67}TiO_3$ (Sintered pellets from Praxair electronics)) with a lens, resulting in a spotsize of 2.3 mm$^2$ (0.9 by 2.5 mm). The beam energy was controlled with a variable attenuator, yielding a fluence at the target of 2 J/cm$^2$. The repetition rate was 1 Hz and the substrate was placed at 5 cm distance directly opposite to the target. Before deposition, the targets were pre-ablated for 2 minutes at 5 Hz to remove any possible surface contamination. After deposition, the PLD chamber was flooded with pure oxygen (typically 100 mbar) and the samples were cooled down by switching of the heater power. Typically, the cooldown required 2 hours. The optimization of the settings for the growth of LSMO is described elsewhere.[6,22]

Atomic resolution STEM-EELS measurements were performed using a probe-corrected TITAN G2 80-300 (FEI) instrument equipped with a GIF quantum spectrometer for electron energy loss spectroscopy (EELS). The effective probe-size during acquisition is approximately equal to 5 Å. Low loss and core-loss spectra are recorded quasi-simultaneously by using the spectrometer in dual EELS mode. The collection and convergence angle are $\alpha$ = 21 mrad and $\beta$ = 25 mrad, respectively. The energy resolution in STEM-EELS was approximately equal to 1.2 eV.

The magnetization measurements were performed with a vibrating sample magnetometer (VSM) (Physical Properties Measurement System (PPMS) by Quantum Design). For each datapoint in the graph, a full hysteresis loop between 240 and -240 kA/m (~3000 Oe) was measured and the saturation magnetization was calculated after a linear background (diamagnetic contribution from the STO substrate) subtraction. The resistivity of the samples was measured in the van der Pauw configuration (PPMS by Quantum Design). In order to obtain ohmic contacts between the aluminum bonding wires and the LSMO layer, gold contacts were deposited on the corners of the sample with the use of a shadow mask.

**Acknowledgements**

We wish to acknowledge the financial support of the Dutch Science Foundation (NWO) and the Dutch Nanotechnology program NanoNed. S.B. acknowledges the financial support from the European Union under the Framework 6 program under a contract for an Integrated Infrastructure Initiative. Reference 026019 ESTEEM. J.V. and G.V.T. acknowledge funding from the European Research Council under the 7th Framework Program (FP7), ERC grant N°246791 - COUNTATOMS. R.E. acknowledges funding from IFOX. We thank Sandra Van Aert for stimulating discussions.

Received: ((will be filled in by the editorial staff))
Revised: ((will be filled in by the editorial staff))
Published online: ((will be filled in by the editorial staff))

# Captions :

**Figure 1.** The LSMO/STO interface configurations. a) $La_{0.67}Sr_{0.33}O$ terminated interface. b) $MnO_2$ terminated interface. c) Interface engineered $TiO_2/La_{0.33}Sr_{0.67}O/MnO_2$ interface I. d) Interface engineered $TiO_2/La_{0.33}Sr_{0.67}O/MnO_2$ interface II. The electrostatic potential due to the polar discontinuity is indicated with a black line (before reconstruction) and a green dotted line (after reconstruction). The numbers indicate the net charge at each layer, after reconstruction.

**Figure 2**. Fabrication of epitaxial LSMO/STO heterostructures. RHEED specular spot intensity oscillations during the growth of (a) non interface engineered ($n = 8$ uc nIE) sample and (b) interface engineered (8 uc IE) sample. (c,d) The corresponding atomic force micrographs of the surface topography of respectively the non interface engineered and interface engineered samples exhibiting the unit cell height terrace steps.

**Figure 3.** EELS analysis of the LSMO/STO heterostructures ($n = 10$) for the (a) non interface engineered (nIE) and (b) interface engineered (IE) samples. Left) Quantitative color map together with a schematic of the sample structure showing the STO (blue), the LSMO (red) and the sub unit cell $La_{0.33}Sr_{0.67}O$ layer (black). Middle). Normalized core-loss signals for La $M_{4,5}$ (red), Mn $L_{2,3}$ (green), Ti $L_{2,3}$ (blue) and O K (black) edges. Right) Comparison with a weighted Gaussian model (light blue) where the sticks indicate the La occupancies used in the model. For the non interface engineered sample clear La (red) diffusion into the STO can be observed, while no La diffusion is present for the interface engineered sample.

**Figure 4.** Temperature dependent (a) magnetization and (b) resisitivity measurements for LSMO/STO heterostructures of various *n* thicknesses in unit cells (uc). Results of samples with (IE) and without interface engineering (nIE) are shown by respectively closed and open symbols.

**Figure 5**. Thickness dependence of the LSMO/STO heterostructures with (closed circles) and without interface engineering (open circles). a) Saturation magnetization at 10K, b) Curie temperature and c) electrical conductivity at 10K.

**Figure 1**

**a) La$_{0.67}$Sr$_{0.33}$O terminated interface**

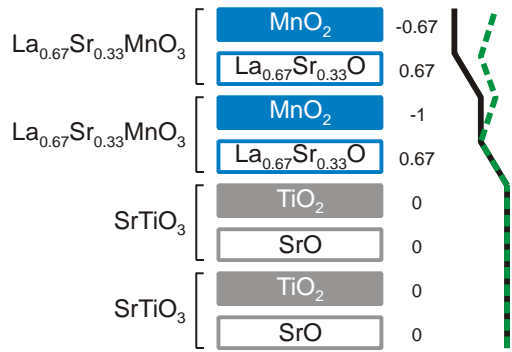

**b) MnO$_2$ terminated interface**

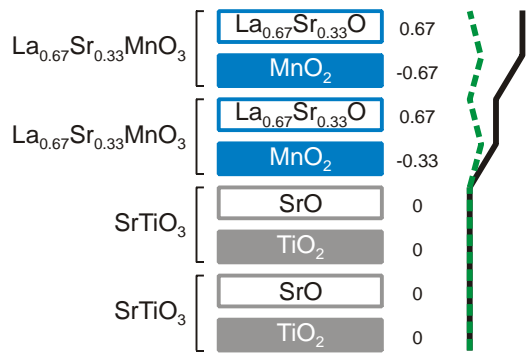

**c) Interface engineered interface I**

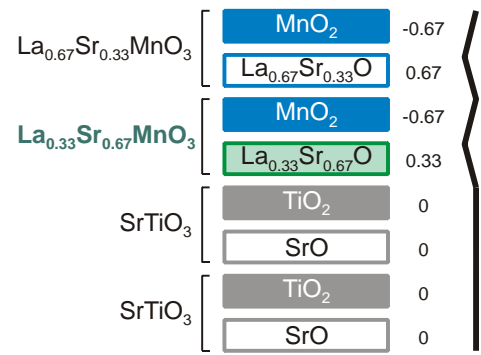

**d) Interface engineered interface II**

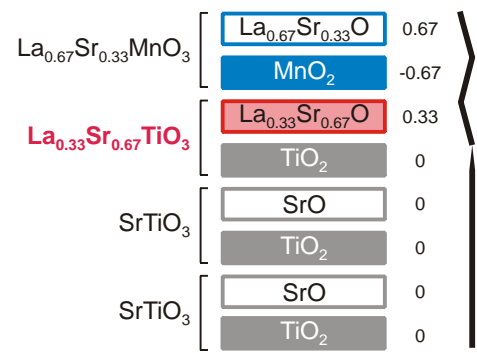

**Figure 2**

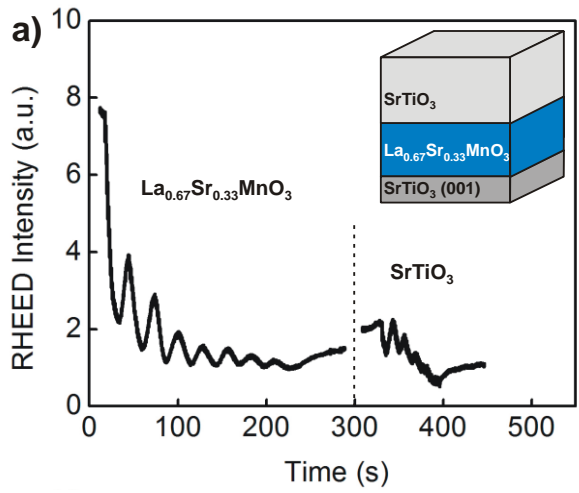

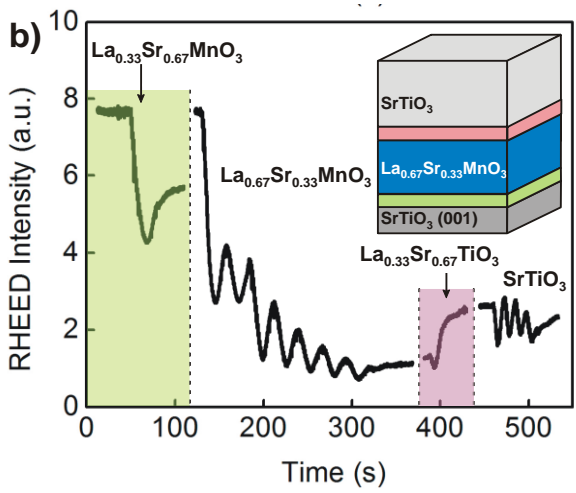

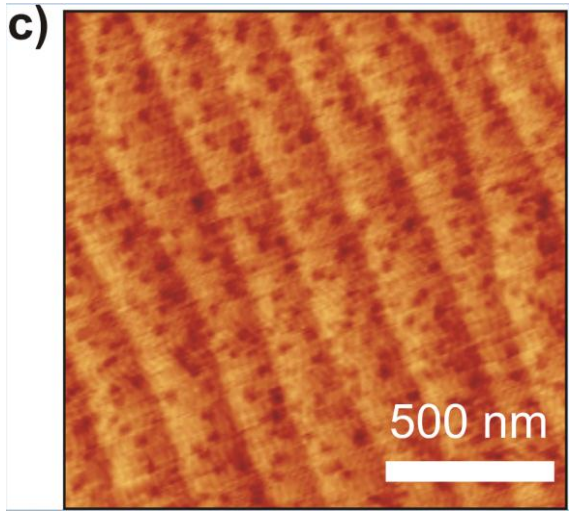

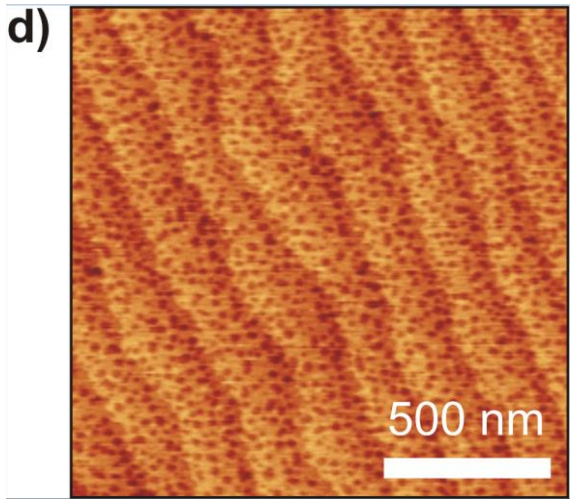

**Figure 3**

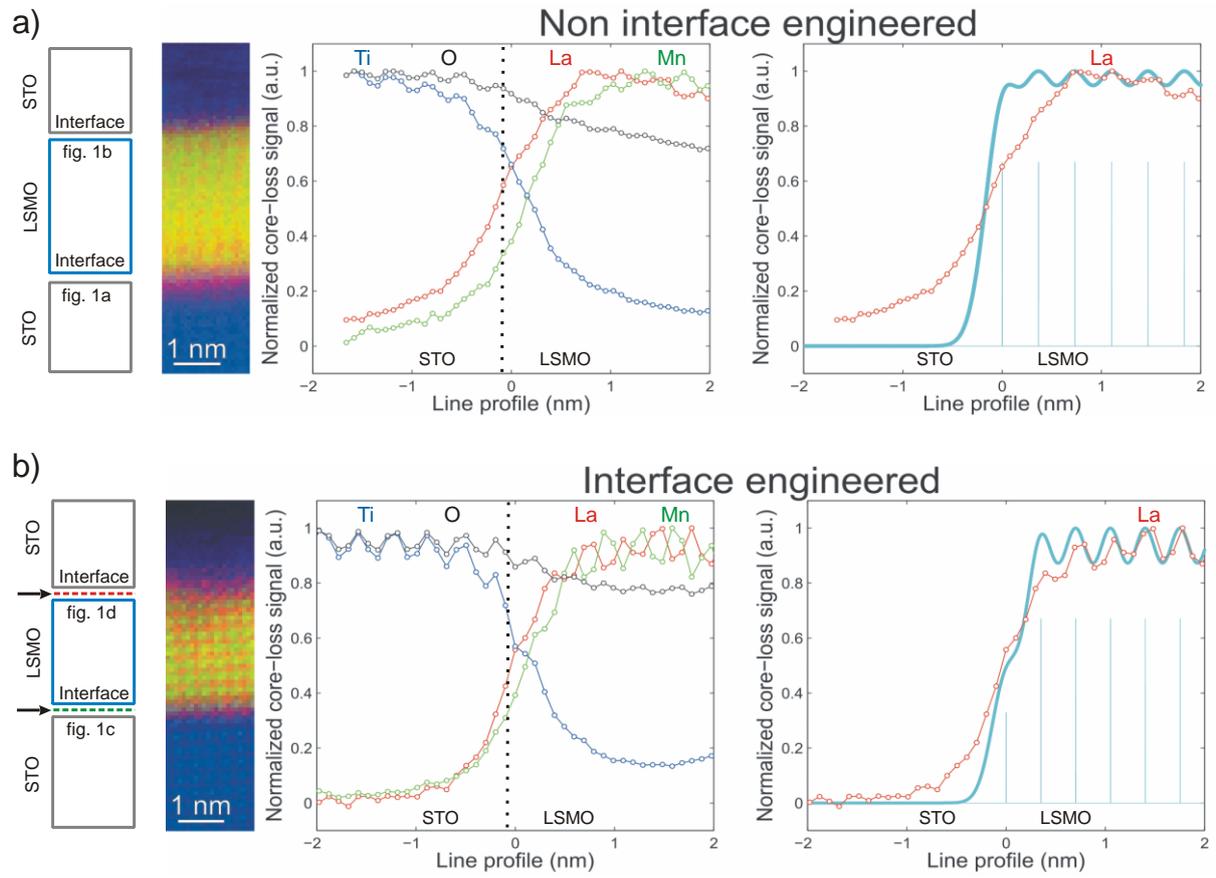

**Figure 4**

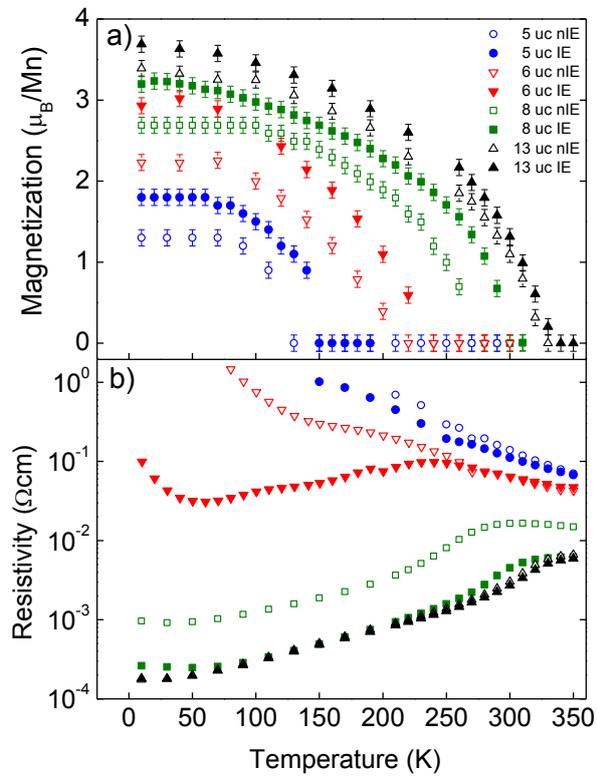

**Figure 5**

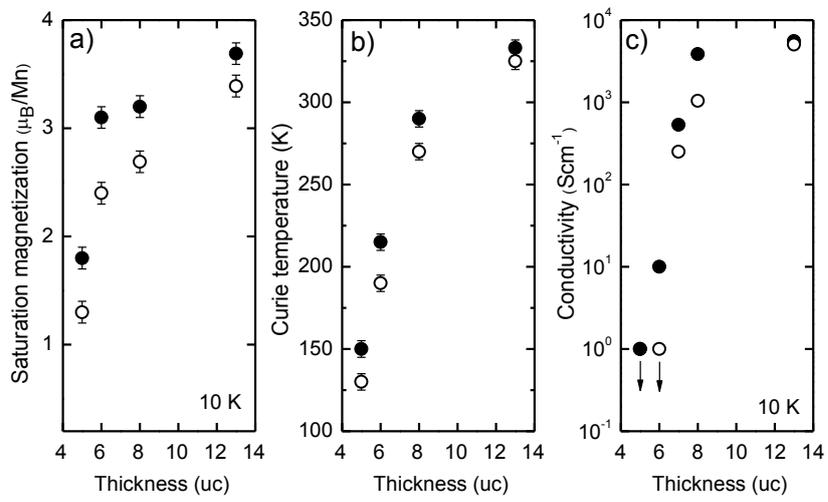